\documentclass[10pt]{article}
\usepackage{epsfig}
\date{}

\title{Chiral phase transition in hadronic matter:\\
the influence of baryon density.}
\author{B.L.~Ioffe\\
\\
{\it\small Institute of Theoretical and Experimental Physics,}\\
{\it\small B.Cheremushkinskaya 25, 117218 Moscow,Russia}}

\newcommand{\be}{\begin{equation}}
\newcommand{\ee}{\end{equation}}
\newcommand{\eq}[1]{(\ref{#1})}

\def\ga{\mathrel{\mathpalette\fun >}}
\def\fun#1#2{\lower3.6pt\vbox{\baselineskip0pt\lineskip.9pt
\ialign{$\mathsurround=0pt#1\hfil##\hfil$\crcr#2\crcr\sim\crcr}}}

\begin{document}

\maketitle

\begin{abstract}A qualitative analysis of the chiral phase
transition in QCD with two massless quarks and non--zero baryon
density is performed. It is assumed that at zero baryonic density,
$\rho=0$, the temperature phase transition is of the second order
and quark condesate $\eta=\langle 0\mid \bar{u} u\mid 0\rangle
=\langle 0 \mid \bar{d}d\mid 0 \rangle $ may be taken as order
parameter of phase  transition. The baryon masses strongly violate
chiral symmetry, $m_B \sim \langle 0 \mid \bar{q}q\mid 0
\rangle^{1/3}$. By supposing, that such specific dependence of
baryon  masses on quark condensate takes place up to phase
transition point, it is shown, that at finite baryon density
$\rho$
 the phase transition becomes of the first order at the
temperature $T=T_{\mathrm{ph}}(\rho)$ for $\rho>0$. At
temperatures $T_{\mathrm{cont}}(\rho) > T > T_{\mathrm{ph}}(\rho)$
there is a mixed phase consisting of the quark phase (stable) and
the hadron phase (unstable). At the temperature $T =
T_{\mathrm{cont}}(\rho)$ the system experiences a continuous
transition to the pure chirally symmetric phase. \end{abstract}


\section{Introduction}

It is well known, that the chiral symmetry is valid in
perturbative quantum chromodynamics (QCD) with massless quarks. It
is expected also, that the chiral symmetry takes place in
full--perturbative and nonperturbative QCD at high temperatures,
($T \ga 200~\mbox{MeV}$), if heavy quarks ($c,b,t$) are ignored.
The chiral symmetry is strongly violated, however, in hadronic
matter, {\it i.e.} in QCD at $T=0$ and low density. What is the
order of phase transition between two phases of QCD with broken
and restored chiral symmetry at variation of temperature and
density is not completely clear now. There are different opinions
about this subject (for a detailed review see
Ref.~\cite{ref:1,ref:wil} and references therein).

In this talk I discuss the phase transitions in QCD with two
massless quarks, $u$ and $d$. Many lattice calculations
\cite{ref:lattice:2nd:eta1,ref:lattice:2nd:noeta1,ref:lattice:2nd:noeta2,ref:lattice:2nd:eta2}
indicate, that at zero chemical potential the phase transition is
of the second order. It will be shown below, that the account of
baryon density drastically changes the situation and the
transition becomes of the first order, and, at high density, the
matter is always in the chirally symmetric phase.

The masses of light $u,d,s$ quarks which enter the QCD Lagrangian,
especially the masses of $u$  and $d$ quarks, from which the usual
(nonstrange)  hadrons are built, are very small, $m_u, m_d < 10$
MeV as compared with characteristic mass scale $M\sim 1$ GeV.
Since in QCD the quark interaction proceeds  through the exchange
of vector gluonic field, then, if light quark masses are
neglected, QCD Lagrangian (its light quark part) is chirally
symmetric, i.e. not only vector, but also axial currents are
conserved and the left and right chirality quark fields are
conserving separately. This chiral symmetry is not realized in
hadronic matter, in the spectrum of hadrons and their low energy
interactions. Indeed, in chirally symmetrical theory the fermion
states must be either  massless or degenerate in parity. It is
evident, that the baryons (particularly, the nucleon) do not
possess such properties. This means, that the chiral symmetry of
QCD Lagrangian is not realized on the spectrum of physical states
and is spontaneously broken. According to Goldstone theorem
spontaneous breaking of symmetry leads to appearance of massless
particles in the spectrum of physical states -- the Goldstone
bosons.

Let us first consider a case of the zero baryonic density and
suppose that the phase transition from chirality violating phase
to the chirality conserving one is of the second order. The second
order phase transition is, generally, characterized by the order
parameter $\eta$. The order parameter is a thermal average of some
operator which may be chosen in various ways. The physical results
are independent on the choice of the order parameter. In QCD the
quark condensate, $\eta = \vert \langle 0 \vert \bar{u} u \vert 0
\rangle \vert = \vert \langle 0 \vert \bar{d} d \vert 0 \rangle
\vert \geq  0 $, may be taken as such parameter. In the phase of
broken chiral symmetry (hadronic phase) the quark condensate is
non-zero and has the normal hadronic scale, $\langle 0\mid
\bar{q}q \mid 0\rangle =(250 MeV)^3$, in the phase of restored
chiral symmetry it is vanishing.

The quark condensate has the desired properties: as it was
demonstrated in the chiral effective theory~\cite{ref:2,ref:3},
$\eta$ decreases with the temperature increasing and an extrapolation
of the curve $\eta(T)$ to the higher temperatures indicates,
that $\eta$ vanishes at $T=T^{(0)}_c \approx 180~\mbox{MeV}$. Here the superscript "0"
indicates that the critical temperature is taken at zero baryon density.
The same conclusion follows  from the lattice
calculations~\cite{ref:lattice:2nd:eta1,ref:lattice:2nd:eta2,ref:Karsch:Review},
where it was also found that the chiral condensate
$\eta$ decreases with increase of the chemical
potential~\cite{ref:mu1,ref:mu2}.

Apply the general theory of the second order phase
transitions~\cite{ref:6} and consider the thermodynamical
potential $\Phi(\eta)$ at the temperature $T$ near $T^{(0)}_c$.
Since $\eta$ is small in this domain, $\Phi(\eta)$ may be
expanded in $\eta$:
\be
\Phi(\eta) = \Phi_0 + \frac{1}{2} A \, \eta^2 + \frac{1}{4} B
\, \eta^4\,, \qquad B > 0\,.
\label{eq:1}
\ee
For a moment we neglect possible derivative terms in the potential.

The terms, proportional to $\eta$ and $\eta^3$ vanish for general
reasons~\cite{ref:6}. In QCD with massless quarks the absence of $\eta$
and $\eta^3$ terms can be proved for any perturbative Feynman
diagrams. At small $t = T - T^{(0)}_c$ the function $A(t)$ is linear in
$t$: $A(t) = a t$, $a>0$. If $t < 0$ the thermodynamical potential
$\Phi(\eta)$ is minimal at $\eta \neq 0$, while at $t > 0$ the
chiral condensate vanishes $\eta = 0$. At small $t$ the $t$-dependence of
the coefficient
$B(t)$ is inessential and may be neglected. The minimum, $\bar{\eta}$, of
the thermodynamical potential can be found from the condition,
$\partial \Phi/\partial \eta = 0$:
\be
\bar{\eta}=
\left\{
\begin{array}{ll}
\sqrt{-at/B}\,, \quad & t < 0\,; \\
0\,, \quad & t > 0\,.
\end{array}
\right.
\label{eq:2}
\ee
It corresponds to the second order phase transition since the potential is quartic in
$\eta$ and -- if the derivative terms are included in the expansion -- the correlation
length becomes infinite at $T=T^{(0)}_c$.

\section{Nucleon mass and quark condensate}

I show now, that the existence of large baryon masses and the
appearance of violating chiral  symmetry quark condensate are
deeply interconnected and even more, that baryon masses arise just
due to  quark condensate. I will use the QCD sum rule method
invented by Shifman, Vainstein and Zakharov [13], in its
applications to baryons [14-17]. (For a review and collection of
relevant original papers see [18]). The idea of the method is that
at virtualities of order $Q^2 \sim 1$ GeV$^2$ the operator product
expansion (OPE) may be used in consideration of hadronic vacuum
correlators. In OPE the nonoperturbative effects reduce to
appearance of vacuum condensates and condensates of the lowest
dimension play the most important role. The perturbative terms are
moderate and do not change the results in essential way,
especially in the cases of chiral symmetry violation, where they
can appear as corrections only.

For definiteness consider the proton mass calculation [14-16].
Introduce the polarization operator

\be
\Pi(p) = i~\int~d^4 x e^{ipx} \langle 0 \vert T { \xi(x),
\bar{\xi} (0) } \vert 0 \rangle \label{(49)} \ee where $\xi(x)$ is
the quark current with proton quantum  numbers and $p^2$ is chosen
to be space-like, \\$p^2 < 0, ~\vert p^2 \vert \sim 1$ $GeV^2$.
The current $\xi$ is the colourless product of three quark fields,
$\xi = \varepsilon^{abc}~q^a q^b q^c,~ q = u, d$, the form of the
current will be specialized below. The general structure of
$\Pi(p)$ is

\be
\Pi(p) = \hat{p} f_1 (p) + f_2(p) \label{(50)} \ee The first
structure, proportional to $\hat{p}$ is conserving chirality,
while the second is chirality violating.

\begin{figure}[tb]
\hspace{25mm} \epsfig{file=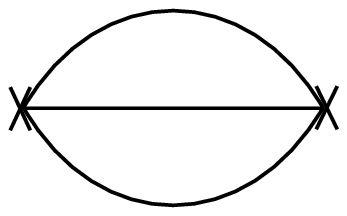} \hspace{39mm}
\epsfig{file=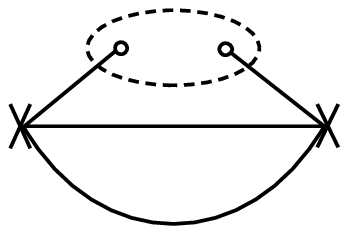}

\vspace{5mm}

\begin{tabular}{p{75mm}p{75mm}}
{\bf Figure 1.}  The bare loop diagram, contributing to chirality
conserving function $f_1(p^2)$: solid lines correspond to quark
propagators, crosses mean the interaction with external currents.
& {\bf Figure 2.}  The diagram, corresponding to chirality
violating dimension 3 operator (quark condensate). The dots,
surrounded by circle mean quarks in the condensate phase. All
other notation is the same as on Fig.1.
\end{tabular}
\end{figure}

For each of the functions $f_i(p^2), ~ i = 1,2$ the OPE can be
written as:

\be
f_i(p^2) = \sum\limits_{n}~ C^{(i)}_n (p^2) \langle 0 \vert
O^{(i)}_n \vert 0 \rangle \label{(51)} \ee where $\langle 0 \vert
O^{(i)}_n \vert 0 \rangle$ are vacuum expectation values (v.e.v)
of various operators (vacuum condensates), $C^{(i)}_n$ are
coefficient functions calculated in QCD. For the first, conserving
chirality structure function $f_i(p^2)$ OPE starts from dimension
zero $(d = 0)$ unit operator. Its contribution is described by the
diagram of Fig.1 and

\be
\hat{p} f_1(p^2) = C_0 \hat{p} p^4 ln [\Lambda^2_u/(-p^2) ] +
polynomial, \label{(52)} \ee where $C_0$ is a constant,
$\Lambda_u$ is the ultraviolet cutoff. The OPE for chirality
violating structure $f_2(p^2)$ starts from $d = 3$ operator, and
its contribution is represented by the diagram of Fig.2:

\be
f_2(p^2) = C_1 p^2 \langle 0 \vert 0 \bar{q} q \vert 0 \rangle ln
\frac{\Lambda^2_u}{(-p^2)} + polynomial \label{(53)} \ee Let us
for a moment restrict ourselves to this first order terms of OPE
and neglect higher order terms (as well the perturbative
corrections).

On the other hand, the polarizaion operator (3) may be expressed
via the characteristics of physical states using the dispersion
relations

\be
f_i(s) = \frac{1}{\pi}~ \int~ \frac{Im f_i
(s^{\prime})}{s^{\prime}+s}~ ds^{\prime} + polynomial,~~ s = -p^2
\label{(54)} \ee

The proton contribution to $Im \Pi(p)$ is equal to

\be
Im \Pi(p) = \pi \langle 0 \vert \xi \vert p \rangle \langle p
\vert \overline{\xi} \vert 0 \rangle \delta(p^2-m^2) = \pi
\lambda^2_N (\hat{p}+m) \delta(p^2-m^2), \label{(55)} \ee where

\be
\langle 0 \vert \xi \vert p \rangle = \lambda_N v(p), \label{(56)}
\ee $\lambda_N$ is a constant, $v(p)$ is the proton spinor and $m$
is the proton mass. Still restricting ourselves to this rough
approximation, we may take equal the calculated in QCD expression
for $\Pi(p)$ (Eq.'s(6),(7)) to its phenomenological representation
Eq.(9). The best way to get rid of unknown polynomial, is to apply
to both sides of the equality the Borel(Laplace) transformation,
defined as

\be
{\cal{B}}_{M^2}f(s) = \lim_{n \to \infty, s \to \infty,\atop
s/n=M^2=Const}~ \frac{s^{n+1}}{n!}~ \Biggl( -\frac{d}{ds} \Biggr
)^n f(s) = \frac{1}{\pi}~ \int\limits^{\infty}_{0}~ ds Im f(s)
e^{-s/M^2} \label{(57)} \ee if $f(s)$ is given by dispersion
relation (8). Notice, that

\be
{\cal{B}}_{M^2} \frac{1}{s^n} = \frac{1}{(n-1)!(M^2)^{n-1}}
\label{(58)} \ee

Owing to the factor $1/(n-1)!$ in (12) the Borel transformation
suppresses the contributions of high order terms in OPE.

Specify now the quark current $\xi(x)$. It is clear from (9) that
proton contribution will dominate in some region of the Borel
parameter $M^2 \sim m^2$ only in the case when both calculated in
QCD functions $f_1$ and $f_2$ are of the same order. This
requirement, together with the requirements of absence of
derivatives  and of renormcovariance fixes the form of current in
unique way (for more details see [14,16]):

\be
\xi(x) = \varepsilon^{abc} (u^a C \gamma_{\mu} u^b) \gamma_{\mu}
\gamma_5 d^c \label{59} \ee where $C$ is the charge conjugation
matrix. With the current $\eta(x)$ (13) the calculations of the
diagrams Fig.2 can be easily performed, the constants $C_0$ and
$C_1$ are determined and after Borel transformation two equations
(sum rules) arise (on the phenomenoloical sides of the sum rules
only proton state is accounted)

\be
M^6 = \tilde{\lambda}^2_N e^{-m^2/M^2} \label{60} \ee

\be
-2 (2 \pi)^2 \langle 0 \vert \bar{q} q \vert 0 \rangle M^4 = m
\tilde{\lambda}^2_N~e^{-m^2/M^2} \label{61} \ee

$$\tilde{\lambda}^2_N = 32\pi^4~\lambda^2_N $$

It can be shown that this rough approximation is valid at $M
\approx m$. Using this value of $M$ and dividing (15) on (14) we
get a simple formula for proton mass [14]:

\be
m = [ - 2 (2 \pi)^2 \langle 0 \vert \bar{q} q \vert 0 \rangle
]^{1/3} \label{(62)} \ee This formula demonstrates the fundamental
fact, that the appearance of the proton mass is caused by
spontaneous violation of chiral invariance: the presence of quark
condensate. (Numerically, (16) gives the experimental value of
proton mass with an accuracy better than 10\%).

A more refined treatment of the problem of the proton mass
calculation was performed: high order terms of OPE were accounted,
as well as excited states in the phenomenological sides of the sum
rules and the stability of the Borel mass dependence was checked.
In the same way, the hyperons, isobar and some resonances masses
were calculated, all in a good agreement with experiment [14,17].
I will not dwell on these results. The main conclusion is: the
origin of baryon masses is in spontaneous violation of chiral
invariance -- the existence of quark condensate in QCD.

\section{The chiral phase transition in the presense of finite\\
baryon density}

I would like to consider here the  influence of baryon density on
the chiral phase transition in hadronic matter. Kogan, Kovner and
Tekin [19] have suggested the idea, that baryons may initiate the
restoration of chiral symmetry, if their density is high -- when
roughly half of the volume is occupied by baryons. The physical
argument in favour of this idea comes from the hypothesis
(supported by calculation in chiral soliton model of nucleon
[20]), that inside the baryon the chiral condensate has the sign
opposite to that in vacuum. This hypothesis  is not proved. Even
more, it is doubtful, that the concept of quark condensate inside
the nucleon can be formulated in a correct way in quantum theory.
But the idea on the strong influence of baryon density on the
chiral phase transition looks very attractive. For this reason no
assumption on the driving mechanism of chiral phase transition at
zero baryon density will be done here. The problem, under
consideration is: how the phase transition changes in the presence
of baryons. The content of my talk  closely follows ref.21.

For completeness it must be mentioned, that the variation of quark
codensate with temperature is not the only source of baryon
effective mass shift. At low $T$ the effective baryon mass shift
arises also due to interaction with pions in thermal bath [22,23].
However, this mass shift, which may be called external (unlike the
internal,  arising from variation of quark condensate) is related
to effective mass, i.e. to propagation of baryon in the matter and
has nothing to do with the properties of the matter as a whole and
the phase transition. (A similar phenomenon takes place in case of
vector mesons, where because of interaction with pions in the
thermal bath the mixing with axial mesons arises [24].)

Consider the case of the finite, but small baryon density $\rho$
(by $\rho$ we mean here the sum of baryon and anti--baryon
densities). For a moment, consider only one type of baryons, {\it
i.e.} the nucleon. The temperature of the phase transition,
$T_{\mathrm{ph}}$, is, in general, dependent on the baryon
density, $T_{\mathrm{ph}} = T_{\mathrm{ph}}(\rho)$, with $
T_{\mathrm{ph}}(\rho=0) \equiv T^{(0)}_c$ At $T <
T_{\mathrm{ph}}(\rho)$ the term, proportional to $E \rho$, where
$E=\sqrt{p^2 + m^2}$ is the baryon energy, must be added to the
thermodynamical potential~\eq{eq:1}. As was shown above the
nucleon mass $m$ (as well as the masses of other baryons) arises
due to the spontaneous violation of the chiral symmetry and is
approximately proportional to the cubic root of the quark
condensate: $m = c \eta^{1/3}$, with $c = (8 \pi^2)^{1/3}$ for a
nucleon. At small temperatures $T$ the baryon contribution to
$\Phi$ is strongly suppressed by the Boltzmann factor $e^{-E/T}$
and is negligible. Below we assume that the proportionality $m
\sim \eta^{1/3}$ is valid in a broad temperature interval.
Arguments in favor of such an assumption are based on the
expectation that the baryon masses vanish at $T =
T_{\mathrm{ph}}(\rho)$ and on the dimensional grounds. Near the
phase transition point $E = \sqrt{p^2 + m^2} \approx p + c^2 \,
\eta^{2/3} / (2 p)$. At $\eta \to 0$ all baryons are accumulating
near zero mass and a summation over all baryons gives us --
instead of eq.~\eq{eq:1} -- the following:
\be
\Phi(\eta, \rho) = \Phi_0 + \frac{1}{2} a t \, \eta^2 +
\frac{1}{4} B \, \eta^4 + C \eta^{2/3} \rho\,, \label{eq:3} \ee
where $C = \sum_i c^2_i/(2 p_i)$. The term $\rho \sum_i p_i$ is
absorbed into $\Phi_0$ since it is independent on the chiral
condensate $\eta$. The typical momenta are of the order of the
temperature, $p_i \sim T$. Thus, Eq.~\eq{eq:3} is valid in the
region $\eta \ll T^3$. In the leading approximation the
coefficient $C$ can be considered as independent on the
temperature at $T \sim T^{(0)}_c$.

Due to the last term in Eq.~\eq{eq:3} the thermodynamical potential
{\it always} have a local minimum at $\eta=0$ since the condensate $\eta$
is always non--negative. At small $t<0$ there also exists a
local minimum at $\eta>0$, which is a solution of the equation:
\be
\frac{\partial \Phi}{\partial \eta} \equiv (a t + B \, \eta^2) \eta +
\frac{2}{3} C \rho \, \eta^{-1/3} = 0\,.
\label{eq:4}
\ee
\begin{figure}[!htb]
\begin{center}
\begin{tabular}{cc}
\includegraphics[angle=-00,scale=1.0,clip=true]{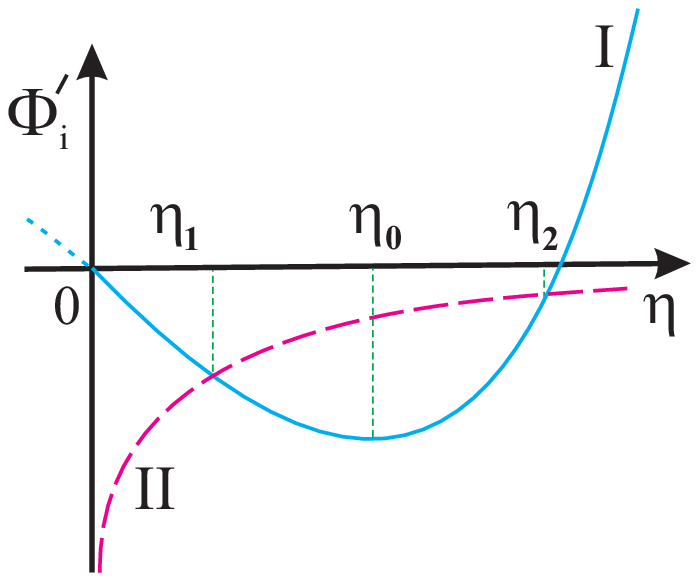} &
\includegraphics[angle=-00,scale=0.8,clip=true]{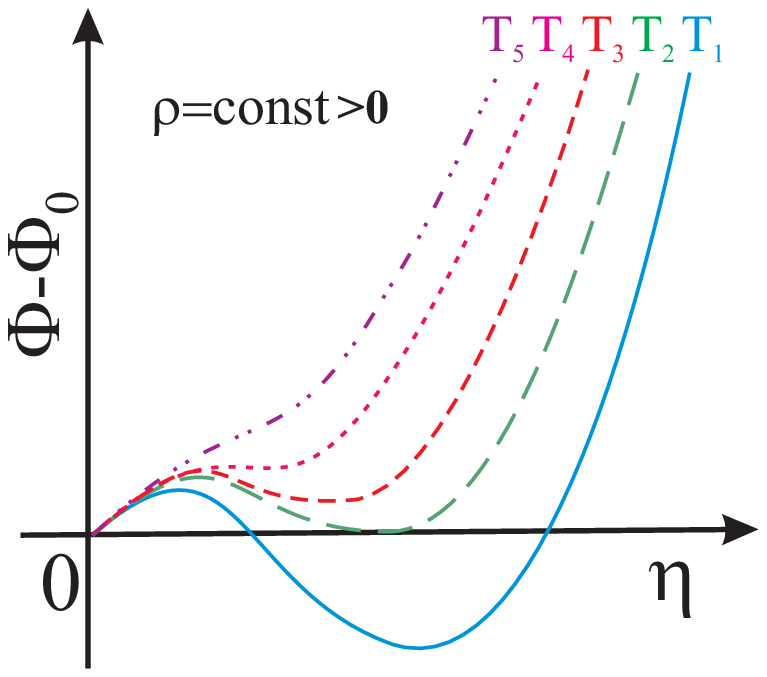} \vspace{1mm} \\
(a) & (b) \\
\end{tabular}
\end{center}

\vspace{5mm}

{\bf Figure 3:} (a) Graphical representation of Eq.~\eq{eq:4}: "I"
is the first term and "II" the second term (with the opposite
sign) in the {\it r.h.s.} of the equation. (b) The thermodynamic
potential~\eq{eq:3} $vs.$ the chiral condensate at a fixed baryon
density $\rho>0$. At low enough temperatures, $T = T_1$, the
system resides in the chirally broken (hadron) phase. The first
order phase transition to the quark phase takes place at
$T_{\mathrm{ph}} = T_2 > T_1$. At somewhat higher temperatures,
$T_3>T_{\mathrm{ph}}$ the system is in a mixed state. The
temperature $T_4 \equiv T_{\mathrm{cont}}$ corresponds to a
continuous transition to the pure quark phase, in which the
thermodynamic potential has the form $T_5$. \label{fig:roots}
\end{figure}

At small enough baryon density $\rho$, Eq.~\eq{eq:4}
 [visualized in Figure~\ref{fig:roots}(a)]
has, in general, two roots, $\eta_1<\eta_0$ and $\eta_2>\eta_0$, where
$\eta_0 = (- a t /3 B)^{1/2}$ is the minimum of the first term in
the right-hand side of Eq.~\eq{eq:4}.
The calculation of the second derivative
$\partial^2 \Phi/\partial \eta^2$ shows that the second root
$\eta_2$ (if exists) corresponds to minimum of $\Phi(\eta)$ and,
therefore, is a local minimum of $\Phi$. The point $\eta = \eta_1$
 corresponds to
a local maximum of the thermodynamical potential since at this point
the second derivative is always non--positive.

The thermodynamical potential $\Phi(\eta, \rho)$ at (fixed) non--zero baryon
density $\rho$ has the form plotted in Figure~\ref{fig:roots}(b). At low
enough temperatures (curve $T_1$) the potential has a global minimum
at $\eta>0$ and system resides in the chirally broken (hadron) phase. As temperature
increases the minima at $\eta=0$ and at $\eta=\bar{\eta}_2>0$ becomes of equal
height (curve $T_2 \equiv T_{\mathrm{ph}}$). At this point the first order
phase transition to the quark phase takes place. At somewhat higher
temperatures, $T=T_3>T_{\mathrm{ph}}$, the $\eta>0$ minimum of the potential
still exist but $\Phi(\eta=0)<\Phi(\bar{\eta}_2)$. This is a mixed phase,
in which the bubbles of the hadron phase may still exist. However, as temperature
increases further, the second minimum disappears (curve
$T_4 \equiv T_{\mathrm{cont}}$). This temperature corresponds to a continuous
transition to the pure quark phase, in which the thermodynamic potential has
the form $T_5$.

Let us calculate the temperature of the phase transition, $T_{\mathrm{ph}}(\rho)$,
at non--zero
baryon density $\rho$. The transition corresponds to the curve $T_2$ in
Figure~\ref{fig:roots}(b), which is defined by the equation
$\Phi(\bar{\eta}_2,\rho)=\Phi(\eta=0,\rho)$, where $\bar{\eta}_2$
is the second root of Eq.~\eq{eq:4} as discussed above. The solution is
\be
T_{\mathrm{ph}}(\rho) = T_c^{(0)} -
\frac{5}{a} {\Biggl(\frac{2 \,C\, \rho}{3}\Biggr)}^{3/5}
\, {\Biggl(\frac{B}{4}\Biggr)}^{2/5}\,,
\label{eq:6}
\ee
and the second minimum of the thermodynamic potential is at
$\bar{\eta}_2 = {[4 a\,(T^{(0)} - T_{\mathrm{ph}}(\rho)) /(5\,B)]}^{1/2}$.

At a temperature slightly higher than $T_{\mathrm{ph}}(\rho)$ the
potential is minimal at $\eta = 0$, but it has also an unstable
minimum at some $\eta>0$. The existence of metastable state is
also a common feature of the first order phase transition ({\it
e.g.}, the overheated liquid in case of liquid--gas system). With
a further increase of the density $\rho$ (at a given temperature)
the intersection of the two curves in Figure 3(a) disappears and
the two curves only touch one another at one point
$\eta=\bar{\eta}_4$. At this temperature a continuous transition
(crossover) takes place. The corresponding potential has the
characteristic form denoted as $T_4$ in Figure 3(a). The
temperature $T_4 \equiv T_{\mathrm{cont}}$ is defined by the
condition that the first~\eq{eq:4} and the second derivatives of
Eq.~\eq{eq:3} vanish:
\be
T_{\mathrm{cont}}(\rho) = T_c^{(0)} - \frac{5}{a}\,
{\Biggl(\frac{2 \,C\, \rho}{9}\Biggr)}^{3/5}
\, {\Biggl(\frac{B}{2}\Biggr)}^{2/5}\,,
\label{eq:cross}
\ee
and the value of the chiral condensate, where the second local minimum of the
potential disappears is given by
$\bar{\eta}_4 = {[2 a (T_{\mathrm{cont}}(\rho) - T^{(0)}_c) /(5\, B)]}^{1/2}$.
At temperatures $T > T_{\mathrm{cont}}(\rho)$ the potential has only
one minimum and the matter is in the state with the restored chiral symmetry.
Thus, in QCD with massless quarks the type of phase transition with the restoration
of the chiral symmetry strongly depends on the value of baryonic density $\rho$.
At a fixed temperature, $T<T^{(0)}_c$, the phase transition happens at
a certain critical density, $\rho_{\mathrm{ph}}$. According to Eq.~\eq{eq:6}
the critical density has a kind of a "universal" dependence on the temperature,
$\rho_{\mathrm{ph}}(T) \propto  [T^{(0)}_c - T]^{5/3}$, the power of which does not
depend on the parameters of the thermodynamic potential, $a$ and $B$.

\begin{figure}[!htb]
\begin{center}
\begin{tabular}{cc}
\includegraphics[angle=-00,scale=1.0,clip=true]{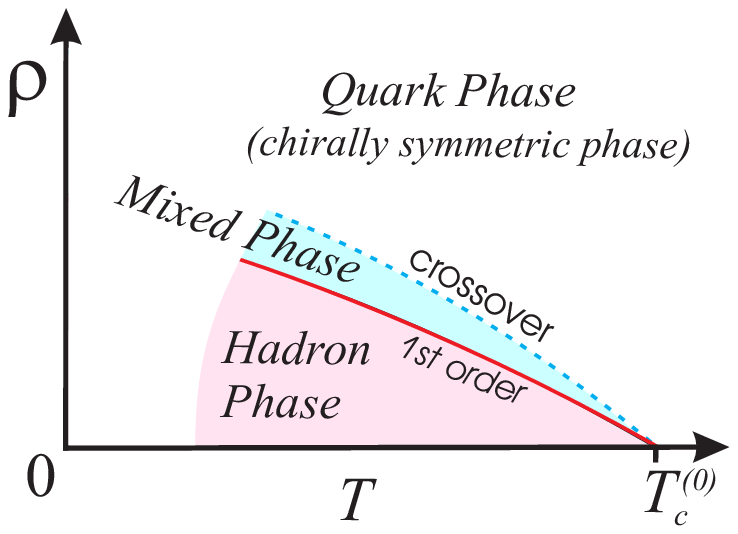} &
\includegraphics[angle=-00,scale=1.0,clip=true]{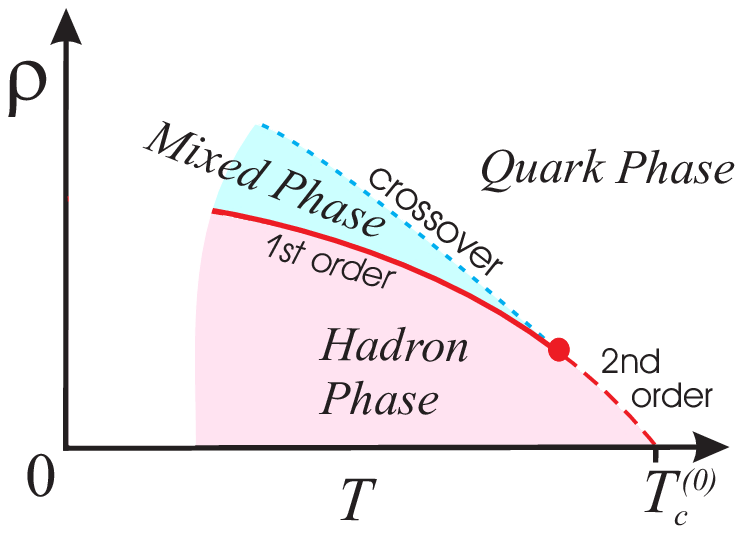} \\
(a) & (b)
\end{tabular}
\end{center}

\vspace{5mm}

{\bf Figure 4}: The qualitative phase diagram at finite baryon
density and temperature based on the analysis (a) without and (b)
with indication of the approximate 2-nd order transition domain.
\label{fig:phase}
\end{figure}

The expected phase diagram is shown qualitatively in Figure 4(a).
This diagram does not contain an end-point which was found in
lattice simulations of the QCD with a finite chemical potential
[25,26]. One may expect that this happens because in our approach
a possible influence of the confinement on the order of the chiral
restoration transition was ignored. Intuitively, it seems that at
low baryon densities such influence is absent indeed: the
deconfinement phenomenon refers to the large quark--anti-quark
separations while the restoration of the of the chiral symmetry
appears due to fluctuations of the gluonic fields in the vicinity
of the quark. However, the confinement phenomenon dictates the
value of the baryon size which can not be ignored at high baryon
densities, when the baryons are overlapping. If the melting of the
baryons happens in the hadron phase depicted in Figure 4(a), then
at high enough density the nature of the transition could be
changed. This may give rise in appearance of the end-point
observed in Ref.[25,26].

The domain where the inequality $|a t| \gg C \rho \eta^{2/3}$,
$\rho \neq 0$ is fulfilled, has specific features. In this domain
the phase transition looks like a smeared second order phase
transition: the specific heat has (approximately) a discontinuity
at the phase transition point, $\Delta C_p = a^2 T_c/2B$. This
statment follows from general theory [12]. At $\mid at\mid \gg
C\rho\eta^{2/3}$ the last term in (17) may be neglected and we
find for the entropy
\be
S=-\frac{\partial \Phi}{\partial T}=S_0 -\frac{1}{2}\frac{\partial
A}{\partial T} \eta^2\ee In the phase above phase transition,
$\eta=0,S=S_0$. Below phase transition
\be
S=S_0 +\frac{a^2}{2B}(T-T_c)\ee The specific heat $C_p=T(\partial
S/\partial T)p$ below the phase transition in the limit $T\to T_c$
is equal
\be
C_p=C_{p_0} +\frac{a^2}{2B}T_c\ee

The correlation length increases as ${(T - T^{(0)}_c)}^{-1/2}$ at
$T - T^{(0)}_c \to 0$. The latter arises if we include the
derivative terms in the effective thermodynamical potential. The
phase diagram with this domain indicated may look as it is shown
in Figure 4(b). Note that the applicability of our considerations
is limited to the region $|T - T_c^{(0)}|/T_c^{(0)} \ll 1$ and low
baryon densities.

In the real QCD the massive heavy quarks (the quarks $c,b,t$)
do not influence on this conclusion, since their concentration
in the vicinity of $T \approx T_c^{(0)} \sim 200~\mbox{MeV}$ is
small. However, the strange quarks, the mass of which $m_s
\approx 150~\mbox{MeV}$ is just of order of expected $T_c^{(0)}$, may
change the situation. This problem deserves further investigation.

This work was supported in part by INTAS grant 2000-587, RFBR
grants 03-02-16209 and by funds from EC to the project ``Study of
Strongly  Interacting Matter'' under contract 2004 No.
R113-CT-2004-506078.


\newpage

\begin{thebibliography}{99}

\bibitem{ref:1} A.V.Smilga,
{\it ``Hot and Dense QCD''}, in Handbook of QCD, Boris Ioffe
Festschrift, ed.by M.Shifman, World Scientific, Singapore 2001,
vol.3, p.2033.

\bibitem{ref:wil} K.~Rajagopal and F.~Wilczek, "The Condensed Matter Physics of QCD",
{\it ibid.}, p. 2061.

\bibitem{ref:lattice:2nd:eta1}
F.~Karsch,
Phys.\ Rev.\ D {\bf 49}, 3791 (1994).

\bibitem{ref:lattice:2nd:noeta1}
F.~Karsch and E.~Laermann, Phys.\ Rev.\ D {\bf 50}, 6954
(1994).

\bibitem{ref:lattice:2nd:noeta2}
S.~Aoki {\it et al.}  [JLQCD Collaboration], Phys.\ Rev.\ D
{\bf 57}, 3910 (1998).

\bibitem{ref:lattice:2nd:eta2}
A.~Ali Khan {\it et al.}  [CP-PACS Collaboration], Phys.\ Rev.\
D {\bf 63}, 034502 (2001).

\bibitem{ref:2} H.Leutwyler, Nucl. Phys. {\bf B} (Proc.Suppl.) {\bf 4},
248 (1988).

\bibitem{ref:3} P.Gerber and H.Leutwyler, Nucl.Phys. B {\bf 321}, 387 (1989).

\bibitem{ref:Karsch:Review}
F.~Karsch, Nucl.\ Phys.\ A {\bf 590}, 367C (1995).

\bibitem{ref:mu1}
J.~B.~Kogut, D.~K.~Sinclair, S.~J.~Hands and S.~E.~Morrison,
Phys.\ Rev.\ D {\bf 64}, 094505 (2001).

\bibitem{ref:mu2} J.~B.~Kogut, D.~Toublan and D.~K.~Sinclair,
Nucl.\ Phys.\ B {\bf 642}, 181 (2002).

\bibitem{ref:6} L.Landau and E.Lifshitz, Statistical Physics, part 1,
Pergamon Press, Oxford, 3-d edition, 1977.


\bibitem{13} M.A.Shifman, A.I.Vainshtein and V.I.Zakharov,
Nucl.Phys. {\bf B147}, 385, 448 (1979).


\bibitem{ref:7} B.L.Ioffe, Nucl.Phys. B{\bf 188}, 317 (1981), E{\bf
191}, 591 (1981).

\bibitem{15} Y.Chung, H.G.Dosch, M.Kremer, and D.Schall,
Nucl.Phys. {\bf B197}, 55 (1982).

\bibitem{16} B.L.Ioffe, Zs.Phys. {\bf C18}, 67 (1983).

\bibitem{ref:8} V.M.Belyaev and B.L.Ioffe, ZhETF {\bf 83}, 876 (1982),
{\bf 84}, 1236 (1983).

\bibitem{18} Vacuum Structure and QCD Sum Rules (Current Physics
-- Sources and Comments, Vol.10), ed. by M.A.Shifman, Amsterdam,
North Holland, 1992.

\bibitem{19} I.I.Kogan, A.Kovner and B.Tekin, Phys.Rev. {\bf D63},
116007 (2001).

\bibitem{20} D.Diakonov and V.Petrov, in Handbook of QCD, Boris Ioffe
Festschrift, ed. by M.Shifman, World Scientific, 2001, v.1, p.359.

\bibitem{21} M.N.Chernodub and B.L.Ioffe, Pis'ma v ZhETF {\bf 79},
 742 (2004), hep-ph/0405042.

\bibitem{aa} H.Leutwyler and A.V.Smilga, Nucl.Phys. {\bf B342}, 302
(1990).

\bibitem{ab} V.L.Eletsky and B.L.Ioffe, Phys.Lett., {\bf B401},
327 (1997).

\bibitem{ab} V.L.Eletsky and B.L.Ioffe, Phys.Rev., {\bf D47},
3083 (1993).


\bibitem{ref:endpoint1}
Z.~Fodor and S.~D.~Katz, JHEP {\bf 0203}, 014 (2002).

\bibitem{ref:endpoint2}
Z.~Fodor and S.~D.~Katz,
hep-lat/0402006.


\end{thebibliography}
\end{document}